# High-harmonic generation in metallic titanium nitride

A. Korobenko[1,*], S. Saha[2], A. T. K. Godfrey[1], M. Gertsvolf[3], A. Yu. Naumov[1], D. M. Villeneuve[1],

A. Boltasseva[2], V. M. Shalaev[2], and P. B. Corkum[1]

**High-harmonic generation is the cornerstone of nonlinear optics. It has been demonstrated in a wide range of crystalline systems including dielectrics, semiconductors, and semi-metals (*1–7*), as well as in gases (*8*, *9*), leaving metals out due to their low damage threshold. Here, we report on the high-harmonic generation in metallic titanium nitride (TiN) films. TiN is a refractory plasmonic metal, known for its high melting temperature and laser damage threshold, with optical properties similar to those of gold (*10*, *11*). We show that TiN can withstand laser pulses with peak intensities as high as 13 TW/cm$^2$, one order of magnitude higher than gold, enabling the emission of intraband harmonics up to photon energies of 11 eV. These harmonics can pave the way for compact and efficient plasmonic devices producing vacuum ultraviolet (VUV) frequency combs. Through numerical calculations and experimental studies, we show that the intensity scaling and angular anisotropy of the emitted VUV radiation stem from the anisotropic conduction band structure of TiN, thus confirming its intraband origin.**

* akoroben@uottawa.ca

In semiconductors, dielectrics, and gases, high-harmonic generation is based on the excitation of free electrons by intense, ultrashort, infrared pulses from bound states to excited states (*2–9*). Emission then arises due to these electrons recombining to their initial state (called interband harmonics in solids). After excitation, high harmonic generation can also occur due to the nonlinearity of their generation probability (often approximated as tunneling) and their anharmonic motion in the conduction band (*3*, *12*); the two effects are lumped together and called intraband harmonics in solids.

While free carrier generation is essential for interband harmonics, the large free-electron concentration of metals suggests that metals will primarily create harmonics by the intraband mechanism. However, the low damage threshold and strong light absorption make these crystalline systems absent until now from the list of materials in which high-harmonic generation was observed.

In this work, we demonstrate high-harmonic emission directly from a metallic compound. To overcome the damage threshold limitation, we use TiN – a refractory metal, known for its high melting point and large laser-damage threshold (*10*, *13*) – together with extremely short laser pulses, to maximize the incident intensity while minimizing the energy deposited in the material.

Experimentally, we determine the multi-shot damage threshold of an epitaxially-grown TiN film. This is the maximal optical intensity it could withstand after being irradiated with multiple (~ 100,000) few-cycle duration, 770 nm laser pulses, before developing irreversible modification. Even though the conduction band is highly populated and the free-carriers lead to a large negative permittivity and high absorption for light (*10*), we found the damage threshold intensity to be as much as an order of magnitude higher compared to some conventional metals studied before. At these strong intensities, the fields caused the metallic electrons to explore highly anharmonic regions of the conduction band, producing harmonics that reached the seventh order (HH7) at a wavelength of 100 nm. To our knowledge, this is the first observation of extreme ultraviolet (XUV) generation in a conductive metal film. This opens possibilities for compact and efficient VUV frequency combs in which the electric field enhancement is achieved inside and around a TiN antenna (*14–16*) to replace the large buildup cavity systems currently used for this purpose (*17*, *18*).

[1] Joint Attosecond Science Laboratory, National Research Council of Canada and University of Ottawa, Ottawa, Ontario K1A0R6, Canada,
[2] Purdue University, School of Electrical & Computer Engineering and Brick Nanotechnology Center, West Lafayette, WIN 47909, USA
[3] National Research Council Canada, Ottawa, ON, K1A 0R6, Canada

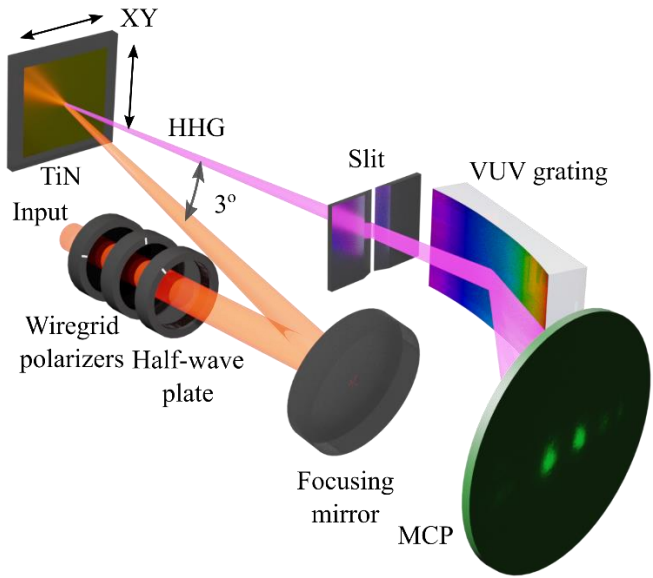

FIG. 1: *Experimental setup. A 2.3-cycle laser pulse (central wavelength 770 nm) was passed through two wire grid polarizers and a half-wave plate. It was focused with a focusing mirror onto the TiN sample inside a vacuum chamber. The sample was mounted on a motorized XY stage, allowing its translation without realigning the optics. The generated high-harmonics radiation (HHG) passed through a slit, diffracted from a curved VUV grating, and reached the imaging MCP detector. The observed VUV spectrum was imaged with a CCD camera.*

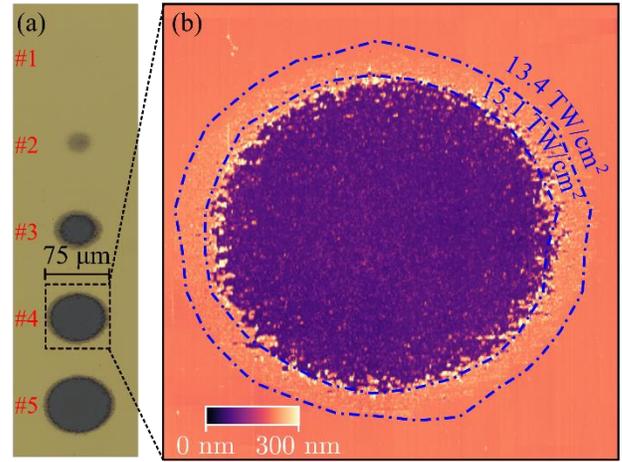

FIG. 2: *(a) Optical microscope image of the irradiated spots on the TiN surface. Numbers 1 through 5 indicate the spots corresponding with the peak field intensities of 12.0, 13.7, 17.1, 20.6, and 24.0 TW/cm$^2$ respectively. We observed modification starting from spot #2, and the film appeared stripped, with the underlying MgO exposed at spots #3, #4 and #5. (b) AFM image of spot #4 reveals a ~150 nm-deep crater, surrounded by a halo of swollen TiN material. The bottom of the crater shows a 40-times increase in surface roughness (17 nm RMS), compared to the unmodified region of the sample (0.4 nm RMS), also showing scattered chunks of material with a characteristic size of 100 nm. The two blue dashed-dotted lines are the contour lines of the independently-measured incident beam profile, corresponding to the peak intensity of 13.4 and 15.1 TW/cm$^2$. These contours set the thresholds for material modification and removal, respectively.*

We perform Density Functional Theory (DFT) calculations to simulate high-harmonic generation in TiN and thereby to confirm that their origin is the intraband motion of the densely populated conduction band electrons. We also predict, and experimentally confirm, the angle dependence of the harmonic conversion efficiency as the laser polarization is rotated with respect to the lattice structure of the solid.

For the experiment (Fig. 1), we used a crystalline TiN film epitaxially grown on a MgO substrate and a 6-fs linearly-polarized pulse, with its polarization vector along the crystal [100] direction. The sample was maintained in a high vacuum.

Fig. 1 shows the layout of the optical setup. To determine the damage threshold, we blocked the laser beam and adjusted its peak intensity with a wire grid polarizer pair. Once the power of the beam was established, it was unblocked, irradiating the film with 60,000 laser pulses. The sample could then be translated by 100 μm to a new spot, and the procedure repeated with a different pulse intensity. After scanning a range of intensities, we removed the sample from the vacuum chamber and inspected it under an optical microscope (Fig. 2 (**a**)) and an atomic force microscope (AFM) (Fig. 2 (**b**)).

Comparing the images with the independently measured incident beam profile, we determined the intensity thresholds to be 13 TW/cm$^2$ and 15 TW/cm$^2$ for the TiN modification and ablation, respectively[4]. This is an order of magnitude higher than the value reported for gold surface for a similar laser pulse (*19*).

TiN's ability to withstand such an extreme laser intensity without damage allowed us to observe high-harmonics, which are otherwise unavailable in metals. The VUV radiation was emitted in the specular direction to the impinging beam. We collected it in a VUV spectrometer (Fig. 1). With the laser polarization fixed along the [100] crystal direction with a half-wave plate, we recorded a set of spectra by varying the laser pulse attenuation with the polarizer (Fig. 3, top panel). Harmonic orders HH5 and HH7 (8.4 and 11.8 eV photon energy, respectively) were observed at the intensities below the TiN damage threshold. They were similar in intensity to the reference harmonics from MgO (measured under the same conditions) (Fig. 3, bottom panel). In addition to HH5 and HH7, harmonic HH9 was also observed from MgO at the intensity range from 10 TW/cm$^2$ to 15 TW/cm$^2$.

---

[4] See the remark on the intensity uncertainty in the Methods

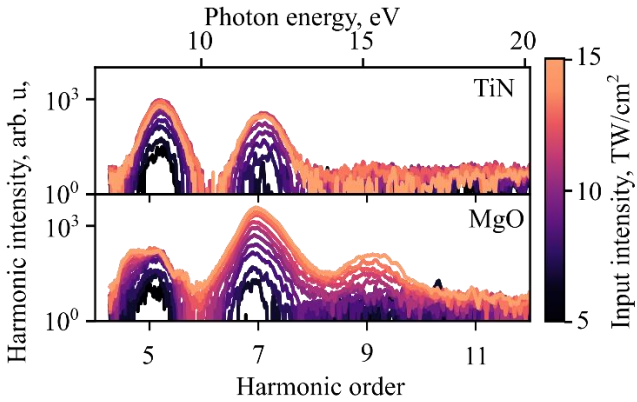

FIG. 3: High harmonic spectra from TiN (top panel) and bare MgO substrate (bottom panel). Lines of different colors represent different input laser peak intensities, as indicated by the color scale on the right. The harmonic intensity was calculated by integrating the MCP image over the vertical dimension. The arbitrary units on the vertical axes are the same on both plots.

Fig. 4(a) summarizes the intensity dependence of the integrated harmonic yield. At the intensity of 13 TW/cm$^2$, marked with the green arrow, the monotonic increase of the TiN harmonics gives way to a decrease as material modification occurs. At intensities greater than 15 TW/cm$^2$, marked with the red arrow, the laser radiation ablates the TiN film, revealing the underlying substrate. As a result, the signal at this intensity is dominated by harmonics generated from the MgO under the thinned-out and stripped TiN film at the bottom of the damage crater, and the HH7 curve is following the seventh harmonic intensity scaling we observe in bare MgO (attenuated due to partial absorption in the leftover TiN).

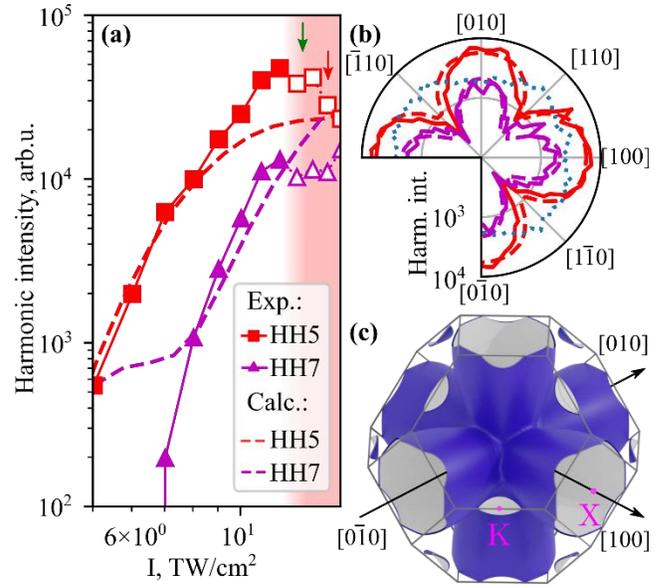

FIG. 4: (**a**) Spectrally integrated intensity of HH5 (squares) and HH7 (triangles), measured as a function of input laser intensity at a constant polarization along the [100] crystallographic direction. Empty markers correspond to intensities above the damage threshold, emphasized by the green arrow. The dashed lines show calculation result. The modelled intensity was scaled up by 40% to reach a better agreement with the experiment. (**b**) HH5 (solid red) and HH7 (solid magenta) intensity, as a function of the laser polarization angle, at a fixed laser peak intensity of 10 TW/cm$^2$. The dashed lines show calculation results. For reference, we plot the angular scan of the HH5 intensity from MgO, measured at the same laser peak intensity, with a blue dotted line. It demonstrates lower anisotropy, and peaks along [110] and symmetrically equivalent directions. (**c**) Highly anisotropic Fermi surface of the TiN conduction band. Grey lines represent the edges of the Brillouin zone of the FCC system.

We confirmed the origin of the harmonics by measuring their angular dependence. We fixed the intensity and scanned the polarization angle relative to the crystal axes, rotating it with a half-wave plate in the (001) crystallographic plane. The results for input intensity of 11 TW/cm$^2$ are shown in Fig. 4 (**b**). Both HH5 and HH7 showed similar anisotropic structure, with the preferable polarization direction along the [100] and symmetrically equivalent crystallographic directions. Comparing the angle dependence of TiN and MgO harmonics, also plotted in Fig. 4 (**b**) with a dotted blue line identifies their distinctive origins.

We attribute the strong anisotropy of the harmonic yields to the anisotropic conduction band structure of the TiN, as reflected in its Fermi surface, shown in Fig. 4 (**c**). The band consists of 6 valleys, centered at X points of the Brillouin zone, elongated in the ΓX direction. This suggests a large difference in the electron dynamics, driven along ΓX and ΓK.

To confirm this hypothesis, we have performed numerical calculations using the semiclassical equations of motion. Unlike in dielectrics and semiconductors, where high harmonic emission is often dominated by electron-hole recollision processes (*3*), in a metal, we expect the main contribution from the intraband current in the highly populated conduction band. We used DFT to retrieve the electronic bands of TiN. In a dielectric, electrons are mostly excited to the conduction band near a single k-point in the Brillouin zone, where the energy gap is the lowest. 1D calculations are, therefore, often sufficient to describe high harmonics. For metals, on the other hand, where the electrons in the conduction band start their trajectories from everywhere in the Brillouin zone, full 3D calculations are necessary. To calculate the harmonic spectra from the band energy $\varepsilon_{\bm{k}}$, we use the Boltzmann equation, that, in the absence of scattering or spatial variation electric field of the laser pulse $\bm{E}(t)$, has a solution $f_{\bm{k}}(t) = f^0_{\bm{k}+e\bm{A}(t)/\hbar}$. Here, $f_{\bm{k}}(t)$ is a time-dependent electron distribution function, $\bm{k}$ is the electron crystal momentum, $e$ is the elementary charge, $\hbar$ is the reduced Planck constant, $\bm{A}(t) = -\int_{-\infty}^{t} dt'\, \bm{E}(t')$ is the vector potential of the laser pulse, $f_{\bm{k}}^0 = \frac{1}{\exp\left(\frac{\varepsilon_{\bm{k}} - E_\text{F}}{k_\text{B} T}\right)+1}$ is the Fermi-Dirac distribution, $E_\text{F}$ is the Fermi energy, $k_\text{B}$ is the Boltzmann's constant and $T$ is the temperature. We then calculate the current density as:

$$\bm{j}(t) = -e \int_{BZ} \frac{d^3 k}{4\pi^3} f_{\bm{k}}(t) \bm{v}_{\bm{k}} \qquad (1)$$

where $\bm{v}_{\bm{k}} = \frac{1}{\hbar} \nabla_{\bm{k}} \varepsilon_{\bm{k}}$ is the electron velocity, $\nabla_{\bm{k}}$ is the gradient operator in reciprocal space, and the integration is carried out over a Brillouin zone.

To reproduce the experimental conditions, an intense, linearly polarized pulse was numerically propagated through the vacuum/TiN interface, using its measured optical constants (see Methods), to find $\bm{A}(t)$ inside. This pulse was then substituted into Eq. (1) to calculate $\bm{j}(t)$. We averaged the resulting current density to account for the intensity profile of the pulse. We then compared the squared amplitude of its Fourier transform with the experiment (Fig. 4 (**a-b**)). While the experimental curves demonstrate monotonic growth, there are oscillations in the calculated ones (Fig. 4 (**a**)). In our experimental intensity range, the oscillations are most pronounced in HH7 (solid magenta line). These oscillations appear to be artifacts due to our neglect of scattering. Indeed, in this model the electron distribution firmly follows $\bm{A}(t)$, exploring non-monotonic and periodic landscape of velocities in the reciprocal space arbitrarily far. Taking scattering, inherent to the real crystals, into account (*20*), eliminates contribution from these unrealistically long trajectories.

We obtained better agreement with the experiment when we calculated the angular dependence of the harmonic yields, plotted in Fig. 4 (**b**). In agreement with the experimental data, the calculations showed four-fold structure, with a substantial increase of the harmonic yield along [100] and symmetrically equivalent directions.

In conclusion, we have observed high-harmonics generation directly from a metallic TiN. We began by measuring the multi-shot damage threshold of TiN illuminated with an intense few-cycle laser pulse. We found TiN to have a damage threshold an order of magnitude higher threshold than common metals, such as gold. This high damage threshold allowed us to observe high harmonics directly from a TiN film, thereby extending the list of high-harmonic generating solids to include metals. The observed spectrum stretched into the technologically important XUV region reaching 11 eV. It is now important to scale the irradiating intensity to the single-shot damage threshold and beyond.

The high harmonics that we measure are intraband harmonics created by conduction band electrons. The obtained harmonic yield is comparable to those generated from a dielectric MgO by the same intensity pulse. In MgO, the harmonic generating electrons are produced by the pulse itself and the harmonics appear to have an interband origin.

Our experiment opens several important technological possibilities. Technologically, since TiN is used to make plasmonic antennas, it will be possible to enhance VUV generation using the field enhancement available with nano-plasmonic antennas (*14–16*). One potentially important application is to produce a compact and stable VUV frequency comb. At present the standard way of generating frequency combs is to increase the amplitude of a weak IR frequency comb field in a power-buildup enhancement cavity (*17*, *18*, *21*), until its intensity is high enough to generate XUV harmonics in a rare gas. We propose to replace the buildup cavity with a TiN nano-plasmonic antenna array and the gas with a dielectric such as MgO.

Another opportunity is to use TiN as an epsilon-near-zero (ENZ (*7*)) material to locally enhance the electromagnetic-field and the nonlinear response. This overcomes the low damage threshold of commonly used transparent conducting oxides such as indium tin oxide (ITO). Since the ENZ wavelength of TiN is around 480 nm (*22*) and can be adjusted, TiN could pave the way to drastically enhanced nonlinear response.

So far, in our experiments, we remained below the multi-shot modification threshold of TiN. Since the single-shot damage thresholds of TiN should be much higher, we will be able to test harmonic conversion efficiency at a much higher intensity by illuminating the sample with a single laser pulse and collecting the generated harmonics spectra. Furthermore, a single-cycle pulse will allow us to far exceed the single-shot damage threshold and still maintain the crystal structure of TiN. Inertially confined (*23*) crystalline metals are an uncharted frontier where the many electrons of a metal can be used to efficiently transfer light from the infrared to the VUV.

At still higher intensity, the pre-existing high carrier concentration of TiN will allow us to study a continuous transition between solid-state high harmonic generation, already linked with gas harmonics, to plasma harmonics, widely studied by the plasma physics community.

## METHODS

### Crystal preparation

A TiN film was deposited using DC magnetron sputtering technique onto a $1 \times 1$ cm$^2$ MgO substrate kept at a temperature of 800°C. A 99.995% pure titanium target of a 2-inch diameter and a DC power of 200 W were used. To ensure high purity of the grown films, the chamber was pumped down to $3 \times 10^{-8}$ Torr before deposition and backfilled to $5 \times 10^{-3}$ Torr with argon during the sputtering process. The throw length of 20 cm ensured a uniform thickness of the grown TiN layer throughout the substrate. After heating, the pressure increased to $1.2 \times 10^{-7}$ Torr. An argon-nitrogen mixture at a rate of 4 sccm/6 sccm was flowed into the chamber. The deposition rate was 2.2 ˚A/min. The surface quality of the grown films was assessed with an atomic force microscope. The films are atomically smooth, with a root-mean-square roughness of 0.4 nm. Their optical properties were characterized via spectroscopic ellipsometry at 50 and 70 degrees for wavelengths of 300 nm to 2000 nm and then fitted with a Drude-Lorentz model, with one Drude oscillator modeling the contribution of the free electrons and two Lorentz oscillators modeling the contribution of the bound electrons.

### Optical setup

We spectrally broadened the 800 nm central wavelength, 1 kHz repetition rate, 1 mJ/pulse energy output of a Ti:Sapph amplifier by passing it through an argon-filled hollow-core fiber. Pulses were then recompressed in a chirped-mirror compressor down to 6 fs FWHM duration, as measured with a dispersion scan technique.

We focused the beam with a 500 mm focal length concave focusing mirror inside a vacuum chamber onto a TiN covered interface of the sample (Fig. 1) at a nearlynormal incidence angle of 1.5˚. The harmonic radiation was emitted from the surface in the specular direction to the incident laser beam, passed through a 300 μm slit of an VUV spectrometer, dispersed by a 300 grooves/mm laminar-type replica diffraction grating (Shimadzu), and an imaging MCP followed by a CCD camera outside the vacuum chamber. We used two wire grid polarizers and a broadband half-wave plate placed outside the chamber to control laser intensity and its polarization. The beam profile at the focal spot was assessed with a CCD camera and found to have a waist radius of 70 μm.

Precise measurement of peak field intensity is diffucult in the case of few-cycle pulses. The values reported in this work were calculated from the measured pulse power, beam profile and temporal characteristics of the pulse. This results in a large uncertainty in the absolute intensity.

### Band structure calculations

Band structure calculations were performed using GPAW package (*24*, *25*), employing a plane-wave basis and PBE exchange-correlation functional, that was found to yield good results in previous DFT studies of TiN (*26*). Having performed the calculations on a rough 16×16×16 *k*-point grid we used Wannier interpolation to interpolate the band energy $\varepsilon_\mathbf{k}$ to a denser 256×256×256 one with wannier90 software (*27*).

The resulting band structure had three energy branches crossing the Fermi level, consistently with previous studies

(*26*, *28*). Two of them had a minimum at the center of the Brillouin zone, Γ, contributing 0.08 and 0.13×10$^{28}$ m$^{-3}$ to the conduction band electron density. The third one, whose Fermi surface is shown in Fig. 4 (**c**), was highly anisotropic and minimized at the X point. Corresponding to the electron density 5.03×10$^{28}$ m$^{-3}$ it was dominant for generating high harmonics.

## Acknowledgements

The work was funded by Air Force Office of Scientific Research (AFOSR) (FA9550-16-1-0109); Canada Research Chairs (CRC); Natural Sciences and Engineering Research Council of Canada (NSERC); National Research Council, Joint Center for Extreme Photonics. We thank David Crane and Ryan Kroeker for their technical support, and are grateful for fruitful discussions with Andre Staudte, Giulio Vampa, Guilmot Ernotte and Marco Taucer.